\documentclass[12pt]{article}
\usepackage{geometry}
\usepackage{makeidx}
\usepackage[T1]{fontenc}
\usepackage[dvipsnames,svgnames,table]{xcolor}
\usepackage{epstopdf}
\usepackage{epsfig}
\usepackage{textcomp}
\usepackage{hyperref}
\usepackage{amsmath}
\usepackage{amssymb}
\usepackage{multirow}
\usepackage{multicol}
\usepackage{booktabs}
\usepackage{caption}
\usepackage{lastpage}
\usepackage{hyperref} 
\usepackage{graphicx}
\usepackage{enumerate}
\usepackage{threeparttable}
\usepackage{float}
\usepackage{array}
\usepackage{cite}
\usepackage{color,soul}
\usepackage{listings}
\usepackage{tikz}
\makeatletter
\renewcommand\section{\@startsection{section}{1}{\z@}%
{-2.5ex \@plus -1ex \@minus -.2ex}%
{2.3ex \@plus.2ex}%
{\normalfont\large\bfseries}}
\renewcommand\subsection{\@startsection{subsection}{1}{\z@}%
{-2.5ex \@plus -1ex \@minus -.2ex}%
{2.3ex \@plus.2ex}%
{\small\bfseries}}

\begin{document}
\bigskip
\title{\textbf{Unveiling neutron and gamma spectrum at Martian surface in presence and absence of hydrogen: A computational study based on GEANT4 simulations}}
\medskip
\author{\small A. Ilker Topuz$^{1}$}
\medskip
\date{\small$^1$Department of Physics, SRM University-AP, Amaravati 522240, India
}
\maketitle
\begin{abstract}
The remote sensing observations of the Martian surface by means of the neutron and gamma-ray spectroscopy indicate the presence of hydrogen in the shallow subsurface of Mars, particularly across several mid- and high-latitude terrains. Additionally, the galactic cosmic ray (GCR) protons incident on the top of the Martian atmosphere induce the generation of the secondary neutrons as well as the secondary gamma rays within the atmosphere and the underlying regolith through hadronic and electromagnetic cascades, a fraction of which, called albedo neutrons and albedo gammas, leak back out from the Martian surface. Motivated by the interactions of the GCR protons with the Martian atmosphere and regolith in the absence and presence of hydrogen, a series of GEANT4 simulations are employed in the current study to unveil the influence of hydrogen on the energy spectra of the secondary and albedo neutrons together with the associated gamma rays at the Martian surface. Again, a discrete 69-bin proton energy spectrum between 0.4 and 115 GeV based on the PAMELA spectrometer is implemented into GEANT4 as performed in a previous study dedicated to the Moon investigation. Subsequently, a geometry that consists of a 12-km-thick Martian atmosphere with a density of $2\times10^{-5}$~g/cm$^3$ along with a CO$_2$-dominated composition and a 2-m-thick single-volume Martian regolith with a bulk density of 3.01~g/cm$^3$ is constructed. The regolith material composition including 45.561 wt.$\%$ SiO$_2$, 0.17 wt.$\%$ TiO$_2$, 3.59 wt.$\%$ Al$_2$O$_3$, 0.37 wt.$\%$ MnO, 14.7 wt.$\%$ FeO, 31.0 wt.$\%$ MgO, 2.88 wt.$\%$ CaO, 0.59 wt.$\%$ Na$_2$O, 0.043 wt.$\%$ K$_2$O, 0.17 wt.$\%$ P$_2$O$_5$, 0.046 wt.$\%$ NiO, and 0.88 wt.$\%$ Cr$_2$O$_3$ is defined in accordance with another study. Then, 0.11 wt.\% hydrogen is introduced to the Martian soil in order to assess the impact of hydrogen on the albedo particles. In the wake of irradiating the Martian atmosphere with a planar vertical PAMELA proton beam, the attenuation of the primary proton spectrum through the atmosphere and the range and energy deposition of protons within the regolith are first obtained. The depth profile of the generated neutrons and gamma rays, along with the kinetic energy spectra of the atmospherically produced neutrons and gammas reaching the surface, is then acquired in the absence and presence of 0.11 wt.\% hydrogen by voxelizing the regolith into 100 cells of 2-cm thickness. Finally, a surface detector is placed at the top of the regolith, and the atmospherically produced neutron and gamma spectra are further compared with the total neutron and gamma spectra recorded at the surface detector in order to assess the relative contribution of the atmosphere and the regolith to the overall radiation field reaching the Martian surface. Then, The energy spectra of the albedo neutrons are obtained in terms of thermal, epithermal, and fast neutrons. From the GEANT4 simulations in this study, it is shown that the Martian regolith, rather than the atmosphere alone, constitutes the dominant source of the neutron and gamma population arriving at the surface, and that the presence of 0.11 wt.\% hydrogen is observable in each energy regime of the albedo neutrons at the Martian surface, while the proton transport and the overall neutron and gamma production depth profile remain comparatively insensitive to hydrogen, thereby providing an indication about the elemental variation of the Martian regolith.
\end{abstract}
\section{Introduction}
\label{Introduction}
The exploration of the Martian surface pertaining to the presence and distribution of hydrogen~\cite{maurice2011mars,martinez2023unfolding} is of particular importance for the future robotic and human exploration missions since the remote sensing techniques like neutron and gamma-ray spectroscopy already hint at the existence of water-equivalent hydrogen (WEH) across several Martian terrains~\cite{maurice2011mars}, which is a picture that has been further refined by the in-situ measurements from the Radiation Assessment Detector (RAD) aboard the Mars Science Laboratory (MSL) rover~\cite{kohler2014measurements,kohler2015measurements}. In this regard, the galactic cosmic ray (GCR) protons that bombard the tenuous Martian atmosphere trigger the generation of the secondary neutrons and gamma rays contingent on the composition and density of the atmosphere and the underlying regolith, and a specific portion of these secondary particles, called albedo neutrons and albedo gammas, leak from the Martian surface~\cite{keating2005model}. Thus, the emission of secondary neutrons and gammas from the Martian surface interacting with the GCR protons provides valuable insights into the composition of the Martian regolith, and the albedo particles that escape from the surface serve as a key indicator of subsurface hydrogen content.

In this study, the influence of hydrogen existing in the Martian regolith on the energy spectra of the albedo neutrons and gammas escaping the surface is investigated by means of the GEANT4 simulations~\cite{agostinelli2003geant4, TopuzGithubMars}. To achieve this aim, a discrete 69-bin proton energy spectrum between 0.4 and 115 GeV based on the PAMELA spectrometer~\cite{adriani2011pamela} is used through a probability grid that is already implemented into GEANT4~\cite{topuz2024effect}. A 12-km-thick Martian atmosphere with a density of $2\times10^{-5}$~g/cm$^3$ is built with a CO$_2$-dominated composition in accordance with the Curiosity rover measurements~\cite{mahaffy2013abundance}, which is underlain by a 2-m-thick single-volume Martian regolith with a bulk density of 3.01~g/cm$^3$ that is consistent with the petrological estimates of the Martian crustal density~\cite{baratoux2014density, yoshizaki2020composition}. The material composition of the regolith, which is built from the oxides of silicon, titanium, aluminum, manganese, iron, magnesium, calcium, sodium, potassium, phosphorus, nickel, and chromium, where SiO$_2$, MgO, and FeO are the dominant constituents, is extracted from the bulk elemental abundance model of the Martian crust~\cite{yoshizaki2020composition}. In the cases where hydrogen is present, 0.11 wt.\% hydrogen is added into the regolith composition by removing an equivalent amount from the constituent oxides, a value chosen to be representative of the WEH abundances inferred for several Martian terrains from orbital neutron spectroscopy~\cite{maurice2011mars}. By injecting a vertical planar proton beam founded on the PAMELA spectrometer at the top of the atmosphere, the attenuation of the primary proton spectrum as well as the range and energy deposited by protons within the regolith is first determined. Then, the generation depth and the kinetic energy spectra of the atmospherically produced neutrons and gammas reaching the surface are obtained in the absence and presence of 0.11 wt.\% hydrogen by voxelizing the regolith. Subsequently, a surface detector is introduced, and the atmospherically produced neutron and gamma spectra are compared with the total neutron and gamma spectra collected at the surface detector, without and with 0.11 wt.\% hydrogen, in order to establish whether the atmosphere or the regolith constitutes the dominant source of the neutron and gamma radiation field reaching the Martian surface. The acquired neutron energy spectra are divided into three energy ranges, namely thermal ($E\leq1$~eV), epithermal ($1~\text{eV}<E\leq1$~keV), and fast ($E>1$~keV)~\cite{NASANeutronranges2026}. This study is organized as follows. In section~\ref{Simulation_setup}, the energy spectrum of the incident GCR protons delivered by the PAMELA spectrometer is exhibited, and the implementation of the 69-bin discrete energy spectrum into GEANT4 is briefly elucidated. Moreover, in the same section, the features of the Martian atmosphere and regolith in the present study are stated in addition to the properties of the GEANT4 simulations. While the current simulation results are shown in conjunction with the proton transport, the atmospherically produced neutron and gamma spectra, the generation depth profile, the atmosphere-versus-regolith contribution to the total surface neutron and gamma population, and the albedo neutron and gamma spectrum by excluding and including 0.11 wt.\% hydrogen in section~\ref{Simulation_outcomes}, section~\ref{Conclusion} finally incorporates the conclusions drawn from the present GEANT4 simulations.

\section{Simulation setup}
\label{Simulation_setup}
\begin{figure}[H]
\begin{center}
\includegraphics[width=12cm]{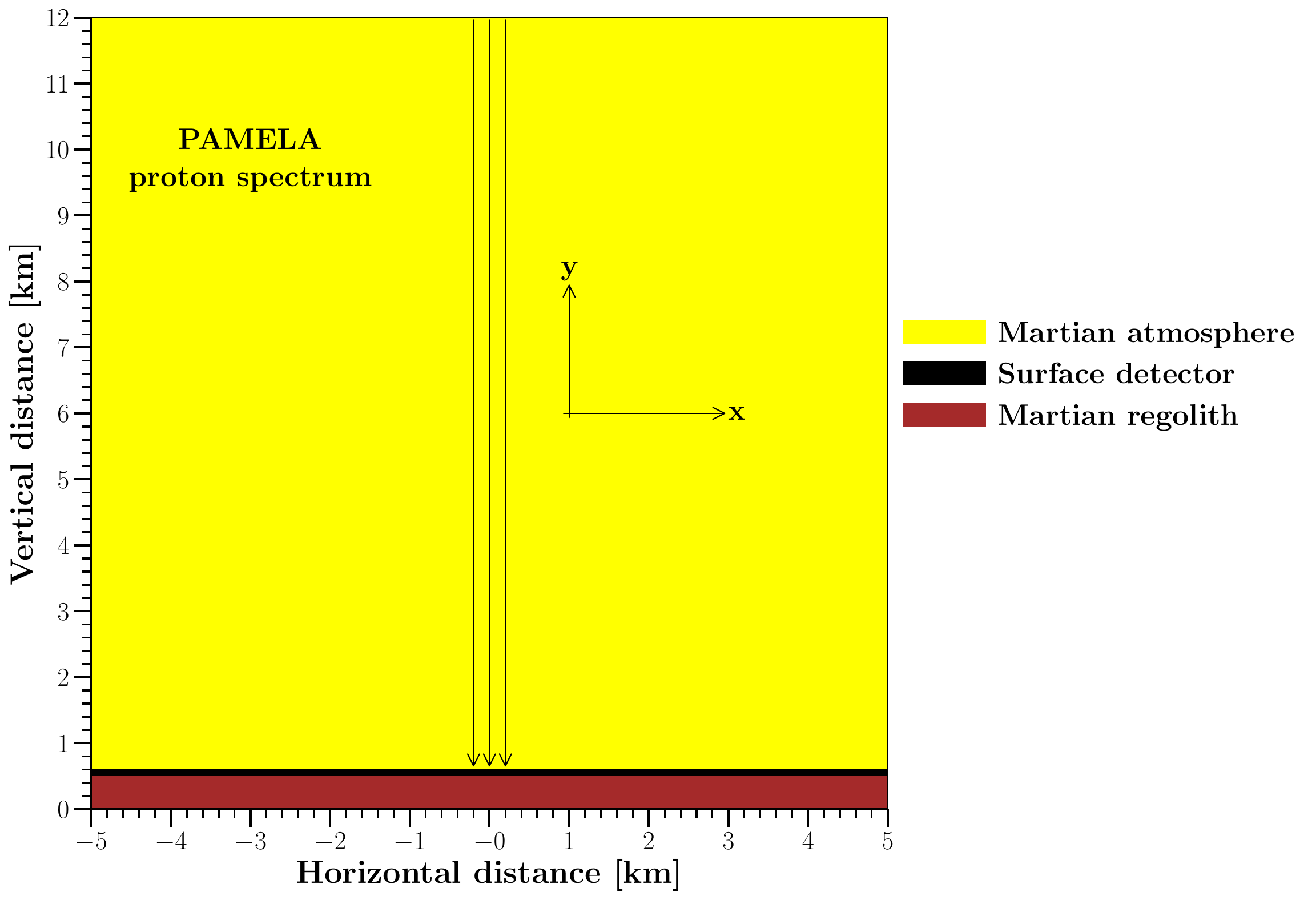}
\caption{Geometrical and structural layout of the 12-km-thick Martian atmosphere and the 2-m-thick Martian regolith in GEANT4 simulations.}
\label{fig:geometry}
\end{center}
\end{figure}
On the first basis, the population and energy spectra of the secondary neutrons and gammas induced by the interaction of GCR protons with the Martian atmosphere and regolith directly depend on structural parameters such as atmospheric column depth, regolith composition, and the resulting density. In this sense, the Martian atmosphere is modeled as a 12-km-thick column of density $2\times10^{-5}$~g/cm$^3$, a value representative of the near-surface Martian atmospheric density~\cite{Seiff1977}, with a CO$_2$-dominated composition consisting of 96.0 wt.\% carbon dioxide, 1.93 wt.\% argon, 1.89 wt.\% nitrogen, and 0.18 wt.\% oxygen in accordance with the Curiosity rover measurements~\cite{mahaffy2013abundance} and summarized in Table~\ref{tab:atmosphere}. 
\begin{table}[H]
\centering
 \captionsetup{width=0.5\columnwidth}
\caption{Elemental/molecular composition and bulk density of the Martian atmosphere used in the present GEANT4 simulations in accordance with Ref.~\cite{mahaffy2013abundance}. The atmospheric density is taken as $2\times10^{-5}$~g/cm$^3$, which is representative of the near-surface value~\cite{Seiff1977}.}
\begin{tabular*}{0.5\columnwidth}{@{\extracolsep{\fill}}*2c}
\toprule
\toprule
Constituent & Mass fraction (wt.\%) \\
\midrule
Carbon dioxide (CO$_2$) & 96.00 \\
Argon (Ar)               & 1.93 \\
Nitrogen (N$_2$)         & 1.89 \\
Oxygen (O$_2$)           & 0.18 \\
\bottomrule
\bottomrule
\end{tabular*}
\label{tab:atmosphere}
\end{table}
Beneath the atmosphere, the Martian regolith is built as a mixture of twelve oxide compounds -- SiO$_2$, TiO$_2$, Al$_2$O$_3$, MnO, FeO, MgO, CaO, Na$_2$O, K$_2$O, P$_2$O$_5$, NiO, and Cr$_2$O$_3$ -- with SiO$_2$, MgO, and FeO as the dominant constituents, in accordance with the bulk elemental abundance model of the Martian crust~\cite{yoshizaki2020composition} and summarized in Table~\ref{tab:regolith}. Furthermore, a single bulky regolith layer of 2-m thickness with a bulk density of 3.01~g/cm$^3$, consistent with petrological estimates of the Martian crustal density~\cite{baratoux2014density}, is taken into consideration to mimic the Martian surface as described in Fig.~\ref{fig:geometry}.
\begin{table}[H]
\centering
 \captionsetup{width=0.5\columnwidth}
\caption{Oxide composition and bulk density of the Martian regolith used in the present GEANT4 simulations in accordance with Ref.~\cite{yoshizaki2020composition}. The regolith bulk density is taken as 3.01~g/cm$^3$~\cite{baratoux2014density}.}
\begin{tabular*}{0.5\columnwidth}{@{\extracolsep{\fill}}*2c}
\toprule
\toprule
Constituent & Mass fraction (wt.\%) \\
\midrule
SiO$_2$  & 45.561 \\
MgO      & 31.00 \\
FeO      & 14.70 \\
Al$_2$O$_3$ & 3.59 \\
CaO      & 2.88 \\
Na$_2$O  & 0.59 \\
Cr$_2$O$_3$ & 0.88 \\
MnO      & 0.37 \\
TiO$_2$  & 0.17 \\
P$_2$O$_5$  & 0.17 \\
NiO      & 0.046 \\
K$_2$O   & 0.043 \\
\bottomrule
\bottomrule
\end{tabular*}
\label{tab:regolith}
\end{table}
As illustrated in Fig.~\ref{fig:geometry}, the Martian atmosphere is defined as a 12-km-thick single volume overlying the 2-m-thick regolith layer. When 0.11 wt.\% hydrogen is introduced into the regolith, the mass percentile of the existing constituent oxides is deducted by the same amount. A pseudo surface detector of 2-mm thick is set atop the regolith at the atmosphere-regolith interface so that the albedo neutrons are collected as illustrated in Fig.~\ref{fig:geometry}. The regolith is voxelized in order to compute the generation depth of the secondary neutrons and gammas, the entire 2-m-thick regolith being divided into 100 cells of 2-cm thickness.

\begin{figure}[H]
\begin{center}
\includegraphics[width=8cm]{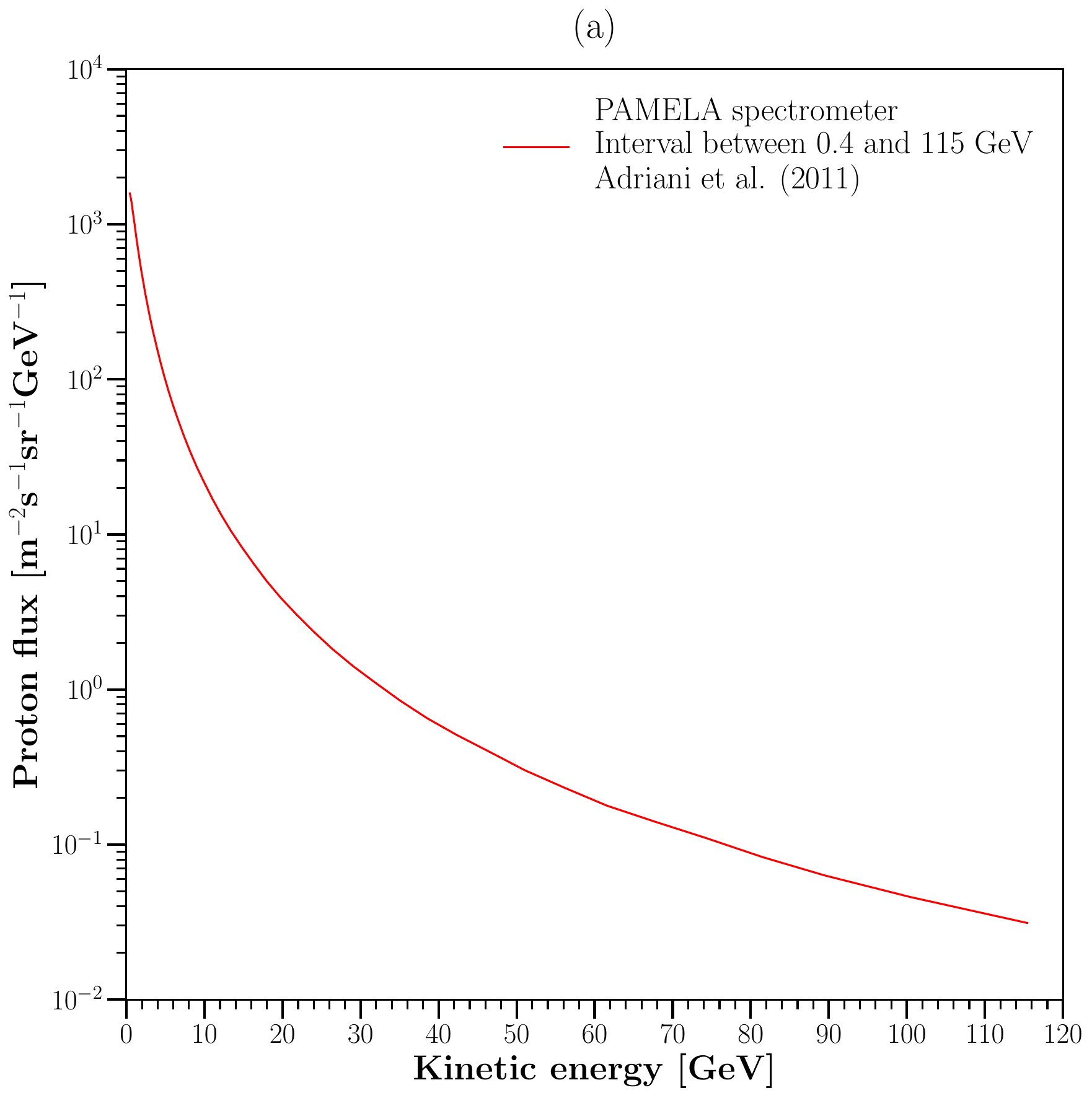}
\includegraphics[width=8cm]{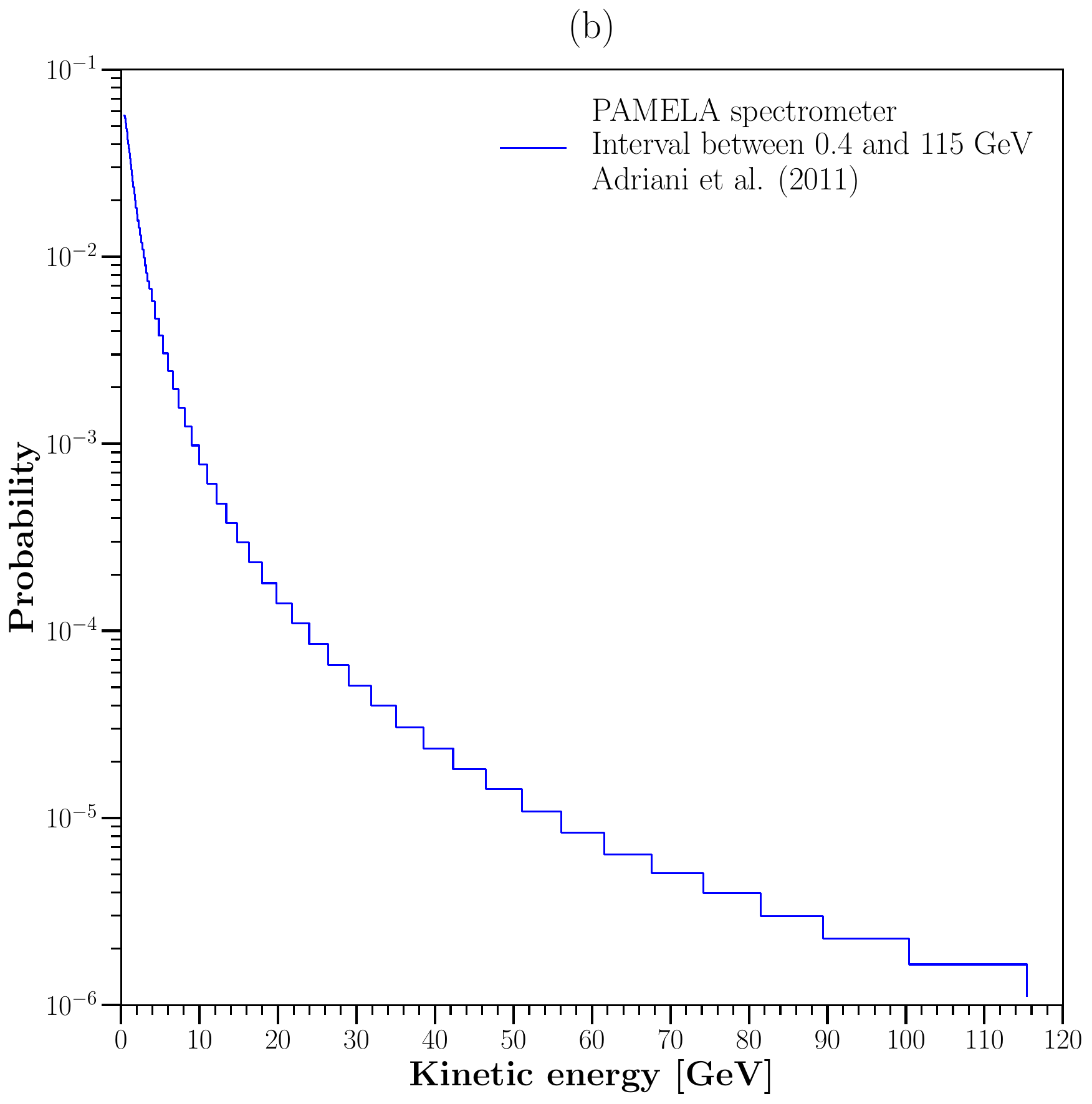}
\caption{PAMELA proton spectrum between 0.4 and 115~GeV (a) particle flux and (b) discrete probabilities.}
\label{fig:pamela}
\end{center}
\end{figure}

In the present study, the Martian atmosphere is irradiated by bombarding it with GCR protons to generate secondary neutrons and gamma rays via the interaction of the GCR protons with the atmosphere and the underlying regolith. To accomplish this objective, among the existing experimental cosmic ray proton spectra is the PAMELA proton spectrum that provides the proton flux values at the discrete energy bins between 0.4 and 1009.4~GeV~\cite{adriani2011pamela}. According to the PAMELA proton spectrum, the cosmic ray proton flux drastically diminishes when the kinetic energy of the protons increases, leading to a proton flux ratio in the order of $10^{-5}$ between 0.4 and 115~GeV as described in Fig.~\ref{fig:pamela}(a). Since the contribution of the particle population after 115~GeV is negligible, a threshold value of 115~GeV is used in the current study.

The tabular proton flux values from the PAMELA spectrometer permit the calculation of the discrete probabilities for each energy bin. In order to implement the PAMELA cosmic ray proton spectrum into GEANT4, the corresponding discrete probability of each energy bin is therefore computed by adding up all the energy bins from 0.4 to 115~GeV and dividing each energy bin by the total sum. Fig.~\ref{fig:pamela}(b) illustrates the variation of the discrete probabilities determined from the PAMELA spectrometer with respect to the kinetic energies. For the purpose of implementing the discrete proton energy spectrum obtained from the PAMELA spectrometer, a strategy to inject the incoming protons is integrated by means of G4ParticleGun. By recalling the unity condition, a grid is built by summing up the discrete probabilities, the interval of which starts with 0 and ends in 1. Thus, each cell in this grid, i.e. the difference between two points on the probability grid, specifies a discrete probability. Then, a random number denoted by $\xi$ between 0 and 1 is generated by using the pre-defined uniform number generator called G4UniformRand(). Finally, this random number is scanned on the probability grid by checking the difference between the grid points, and the particular discrete energy is assigned when the random number matches with the associated cell.
Regarding the simulation features in GEANT4, a planar vertical proton beam based on the PAMELA spectrometer is directed toward the center of the Martian atmosphere and injected along the $y$-axis as indicated in Fig.~\ref{fig:geometry}. In every simulation, the number of incident protons is $10^{5}$, owing to the larger geometry and the additional attenuation introduced by the atmospheric column relative to an airless body. All the materials in the current study are defined in accordance with the GEANT4/NIST material database, and the reference physics list used in these simulations is FTFP\_BERT\_HP. The secondary neutron and gamma tracking inside the voxelized regolith or at the surface detector is maintained by G4Step, and the tracking information about the secondary and albedo particles is post-processed via Python.
\section{Simulation outcomes}
\label{Simulation_outcomes}
\begin{figure}[H]
\begin{center}
\includegraphics[width=7.65cm]{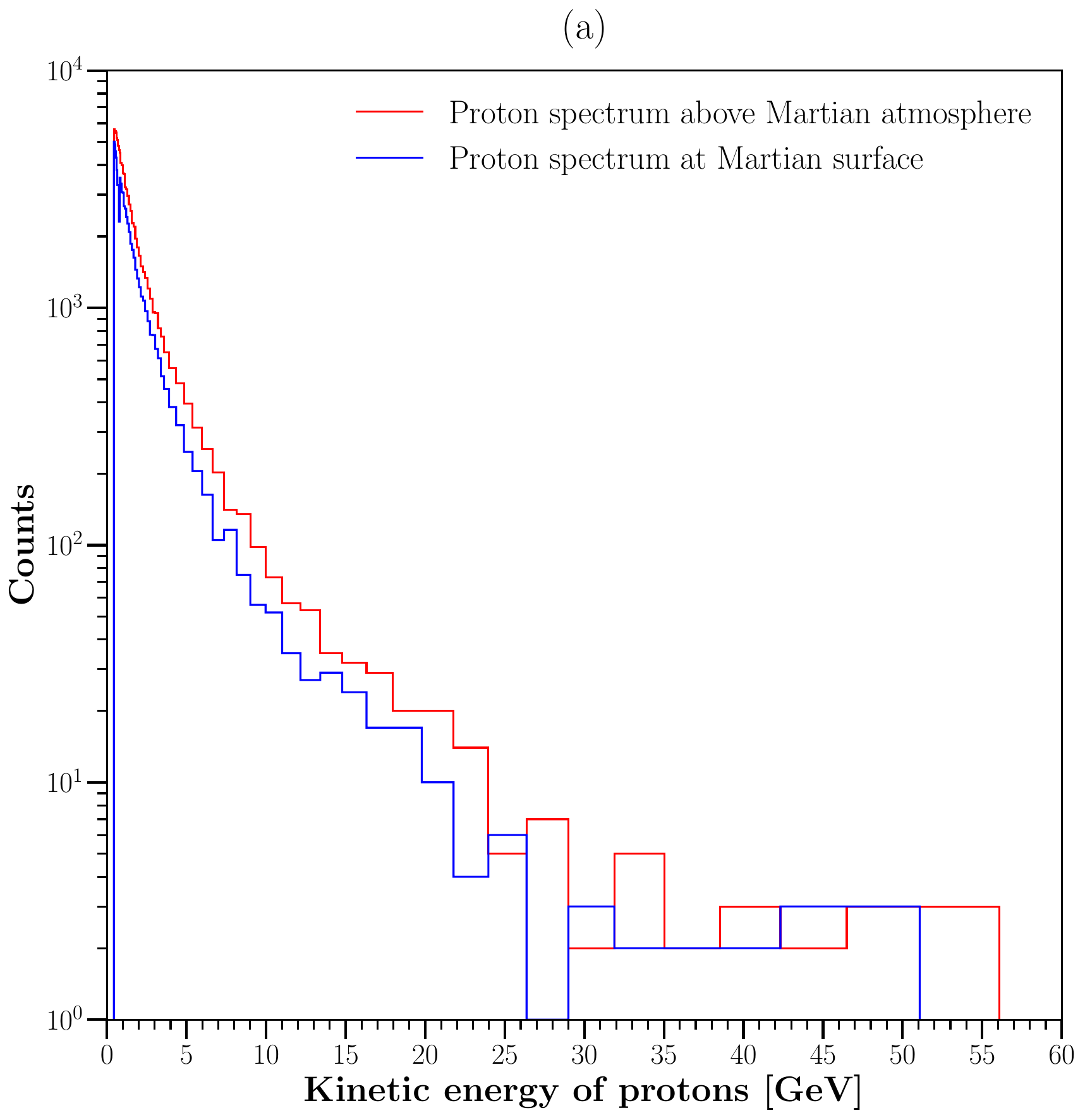}
\includegraphics[width=7.65cm]{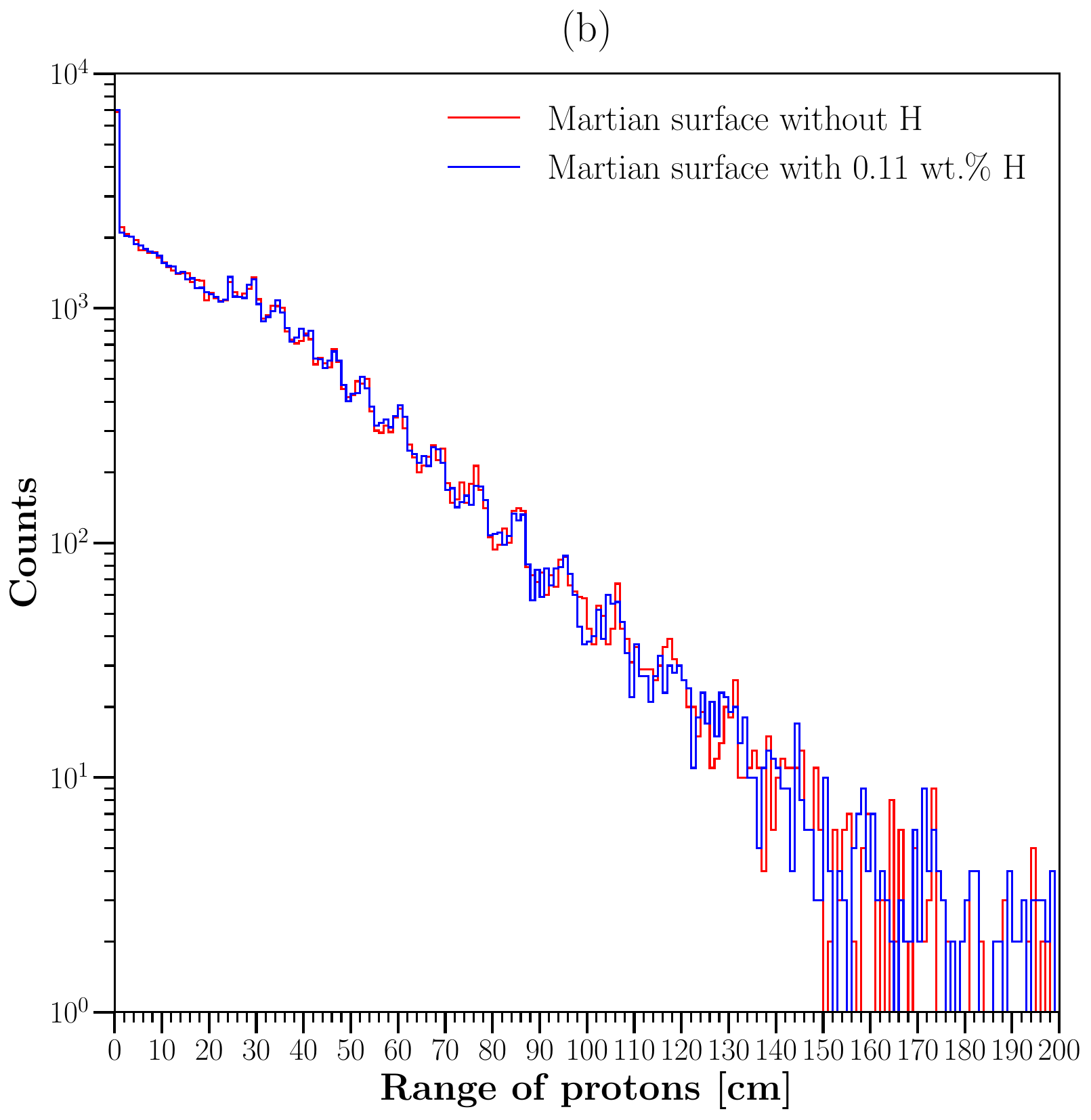}
\includegraphics[width=7.65cm]{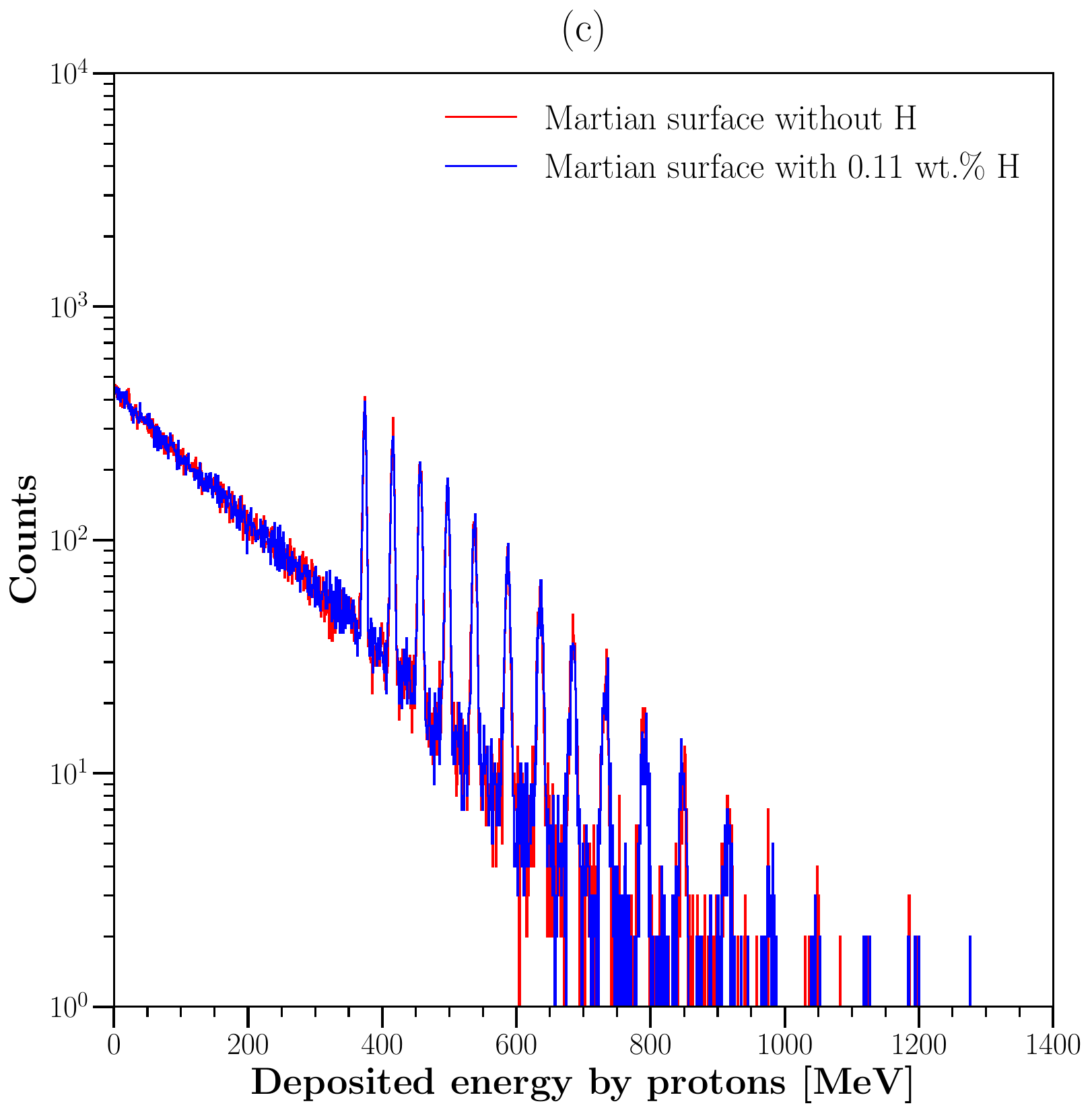}
\caption{Proton transport through the Martian atmosphere and regolith (a) kinetic energy spectrum of protons above the Martian atmosphere compared with the spectrum at the Martian surface, (b) range of protons within the Martian regolith without and with 0.11 wt.\% hydrogen, and (c) energy deposited by protons within the Martian regolith without and with 0.11 wt.\% hydrogen.}
\label{fig:proton_transport}
\end{center}
\end{figure}
The attenuation of the primary GCR proton spectrum through the Martian atmosphere as well as the range and the energy deposited by protons within the regolith is first obtained by comparing the regolith in the absence and presence of 0.11 wt.\% hydrogen. Fig.~\ref{fig:proton_transport}(a) presents the kinetic energy distribution of the primary proton spectrum above the Martian atmosphere together with the distribution of the same protons after traversing the 12-km atmospheric column to reach the surface. As shown in Fig.~\ref{fig:proton_transport}(a), the atmosphere produces a discernible softening of the proton spectrum, with the surface-arriving distribution depleted relative to the incident spectrum and its high-energy tail truncated earlier, on account of ionization energy loss and inelastic nuclear interactions suffered by the protons while traversing the atmospheric column. On the other hand, as illustrated in Figs.~\ref{fig:proton_transport}(b) and~\ref{fig:proton_transport}(c), introducing 0.11 wt.\% hydrogen into the regolith has minimal impact on either the range distribution or the energy-deposition spectrum of the protons, both of which retain the same overall shape without and with hydrogen. It is therefore concluded that, at the trace hydrogen abundance considered here, the direct ionization and nuclear-interaction behavior of the protons is governed almost entirely by the dominant heavier elements of the regolith.

\begin{figure}[H]
\begin{center}
\includegraphics[width=8cm]{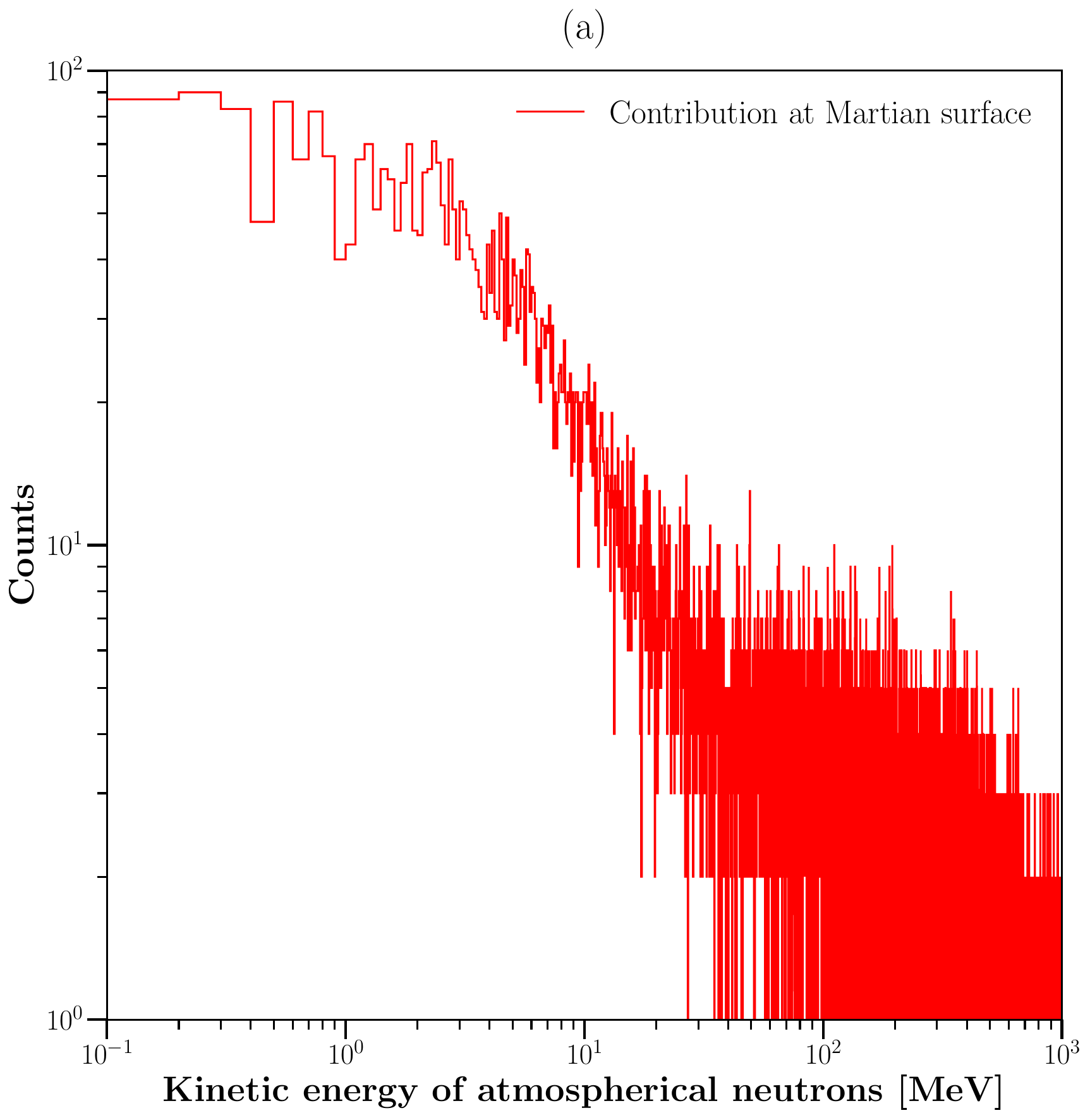}
\includegraphics[width=8cm]{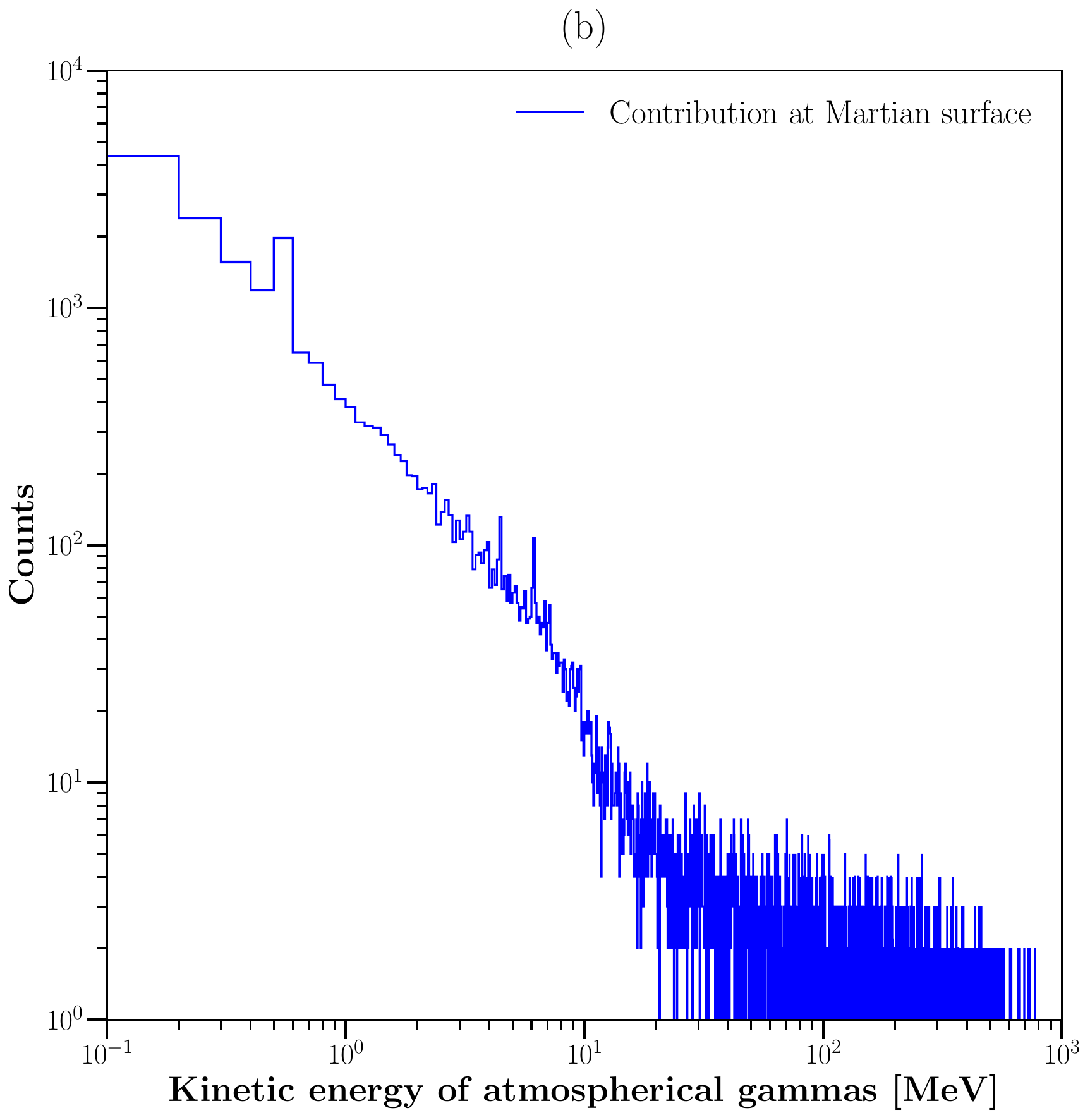}
\caption{Kinetic energy spectra of (a) neutrons and (b) gamma rays produced within the Martian atmosphere and arriving at the surface detector.}
\label{fig:atm_secondaries}
\end{center}
\end{figure}

\begin{figure}[H]
\begin{center}
\includegraphics[width=8cm]{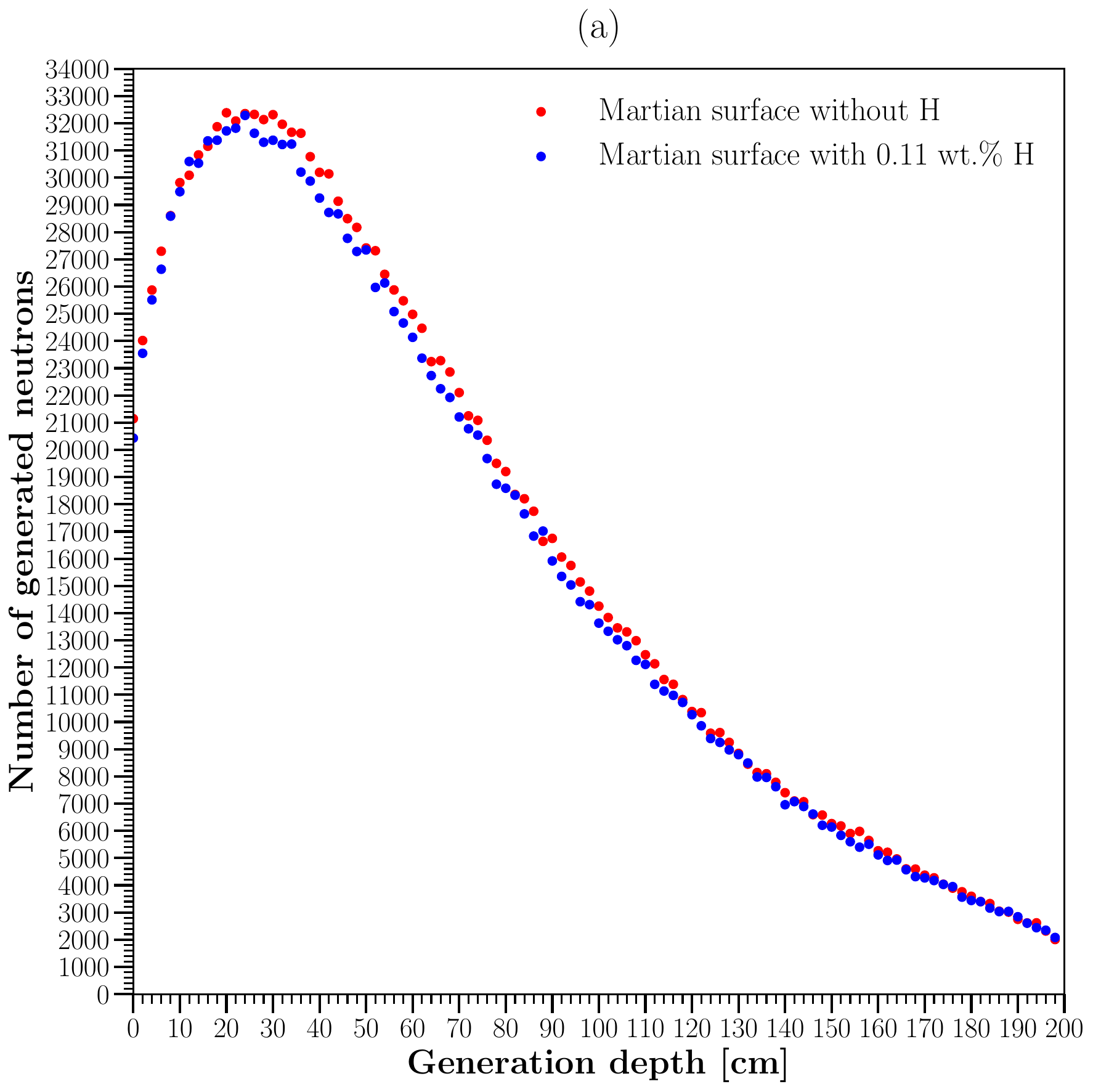}
\includegraphics[width=8cm]{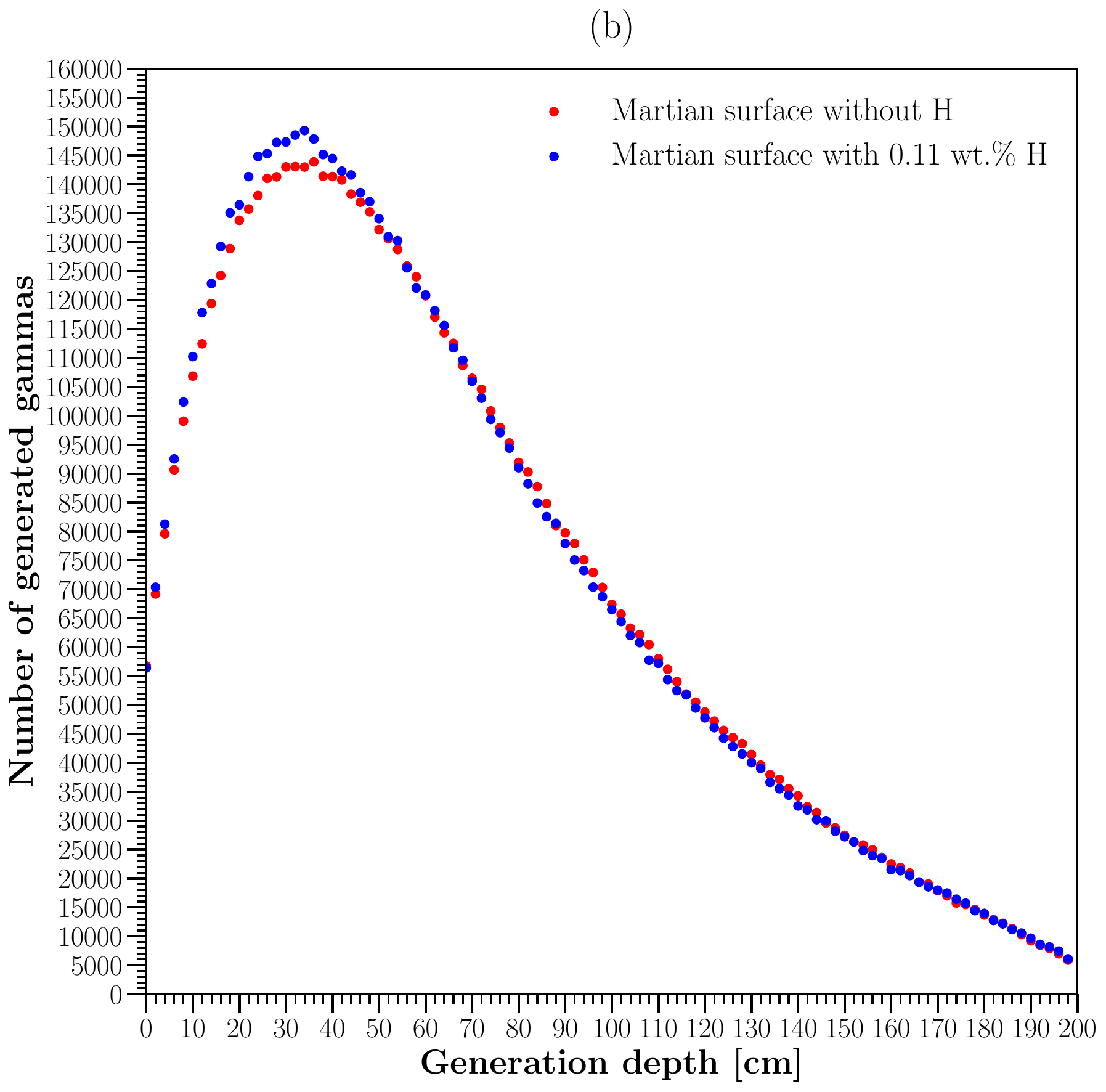}
\caption{Generation depth profile of secondary (a) neutrons and (b) gammas within the Martian regolith, without and with 0.11 wt.\% hydrogen.}
\label{fig:generation_depth}
\end{center}
\end{figure}

\begin{figure}[H]
\begin{center}
\includegraphics[width=8cm]{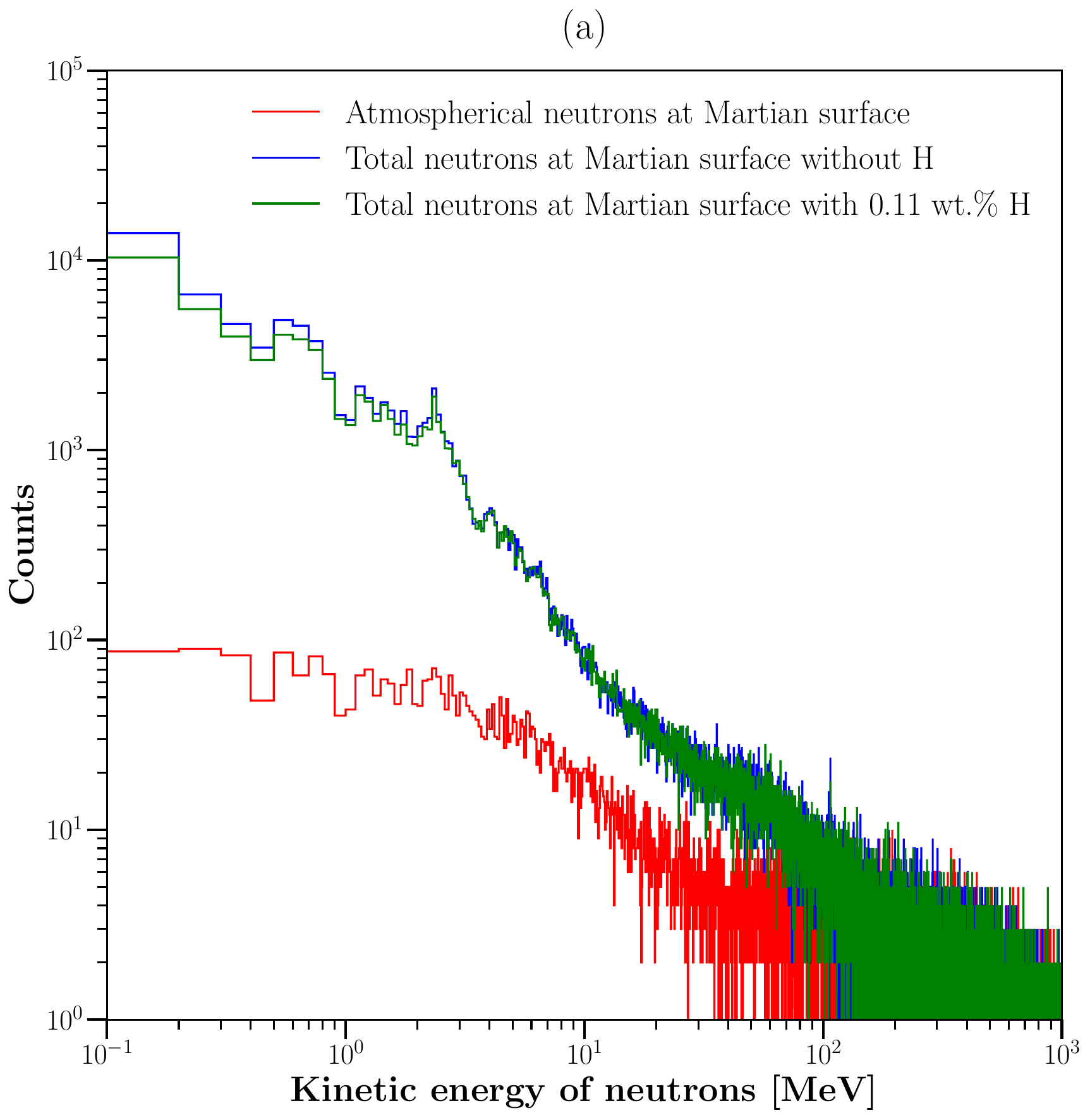}
\includegraphics[width=8cm]{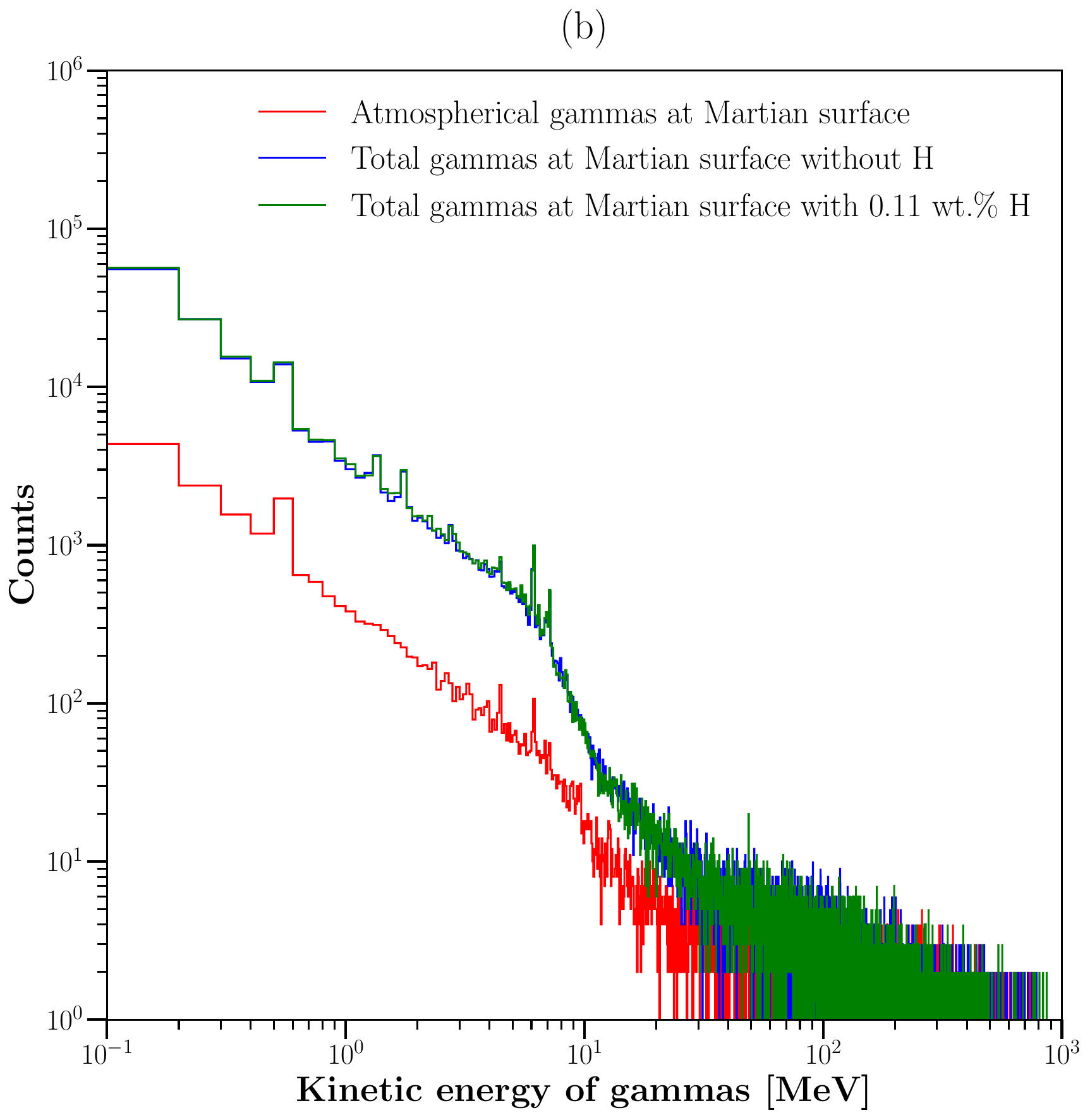}
\caption{Kinetic energy spectra of (a) neutrons and (b) gamma rays at the Martian surface, comparing the atmospherically produced contribution reaching the surface detector prior to any interaction with the regolith with the total neutron and gamma population recorded at the surface detector after accounting for regolith interactions, without and with 0.11 wt.\% hydrogen.}
\label{fig:surface_neutron_gamma}
\end{center}
\end{figure}

\begin{figure}[H]
\begin{center}
\includegraphics[width=8cm]{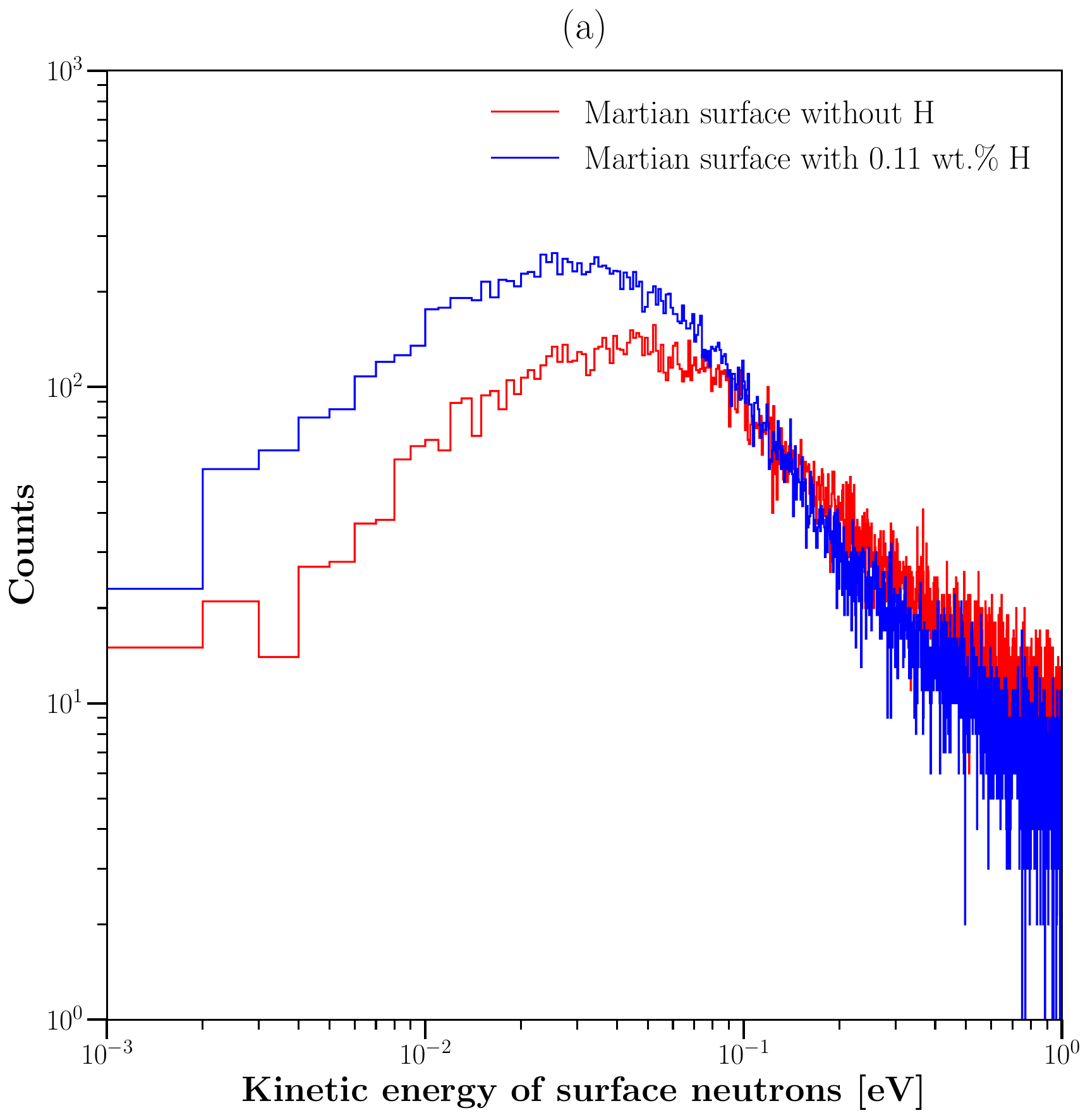}
\includegraphics[width=8cm]{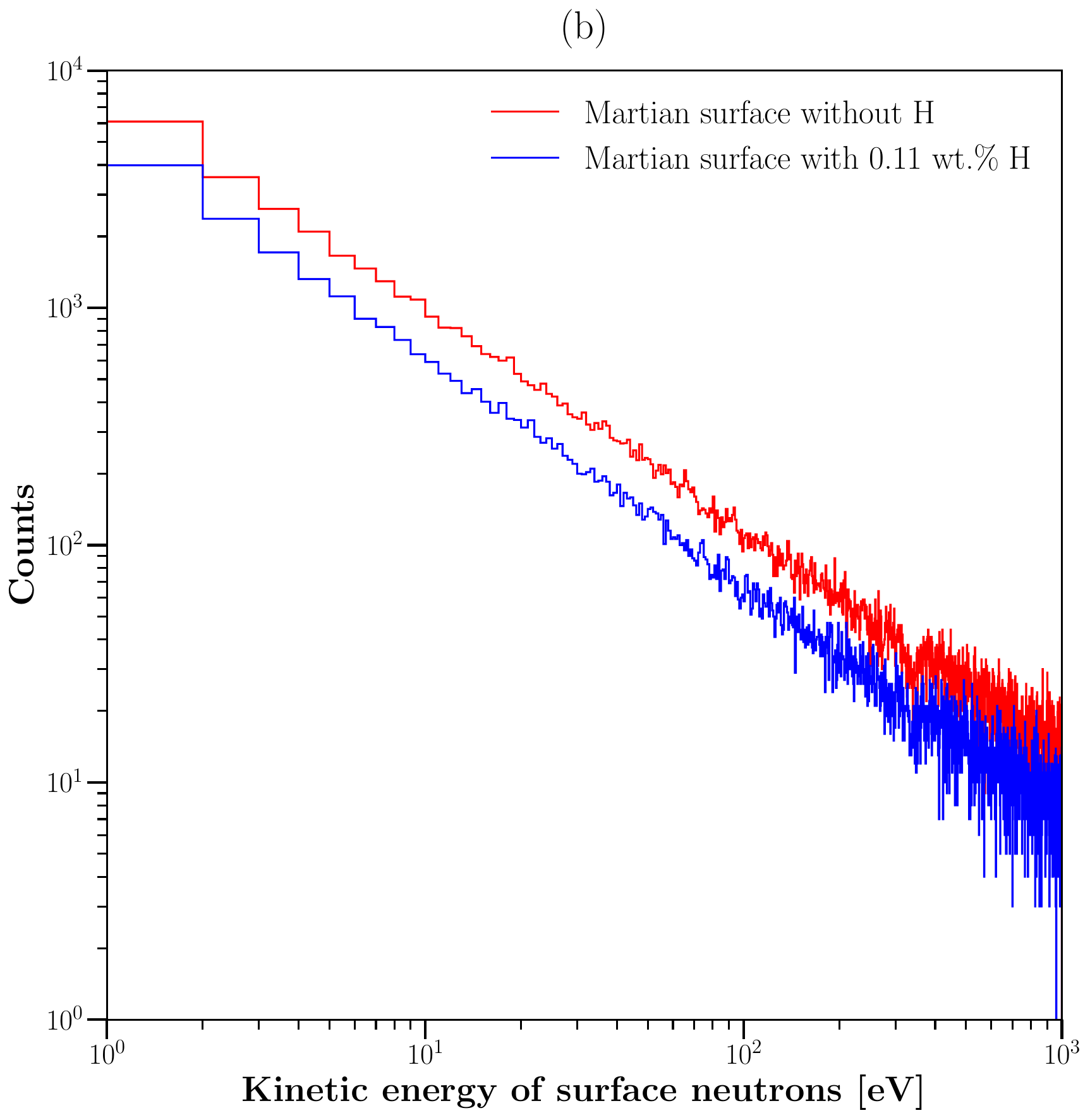}
\includegraphics[width=8cm]{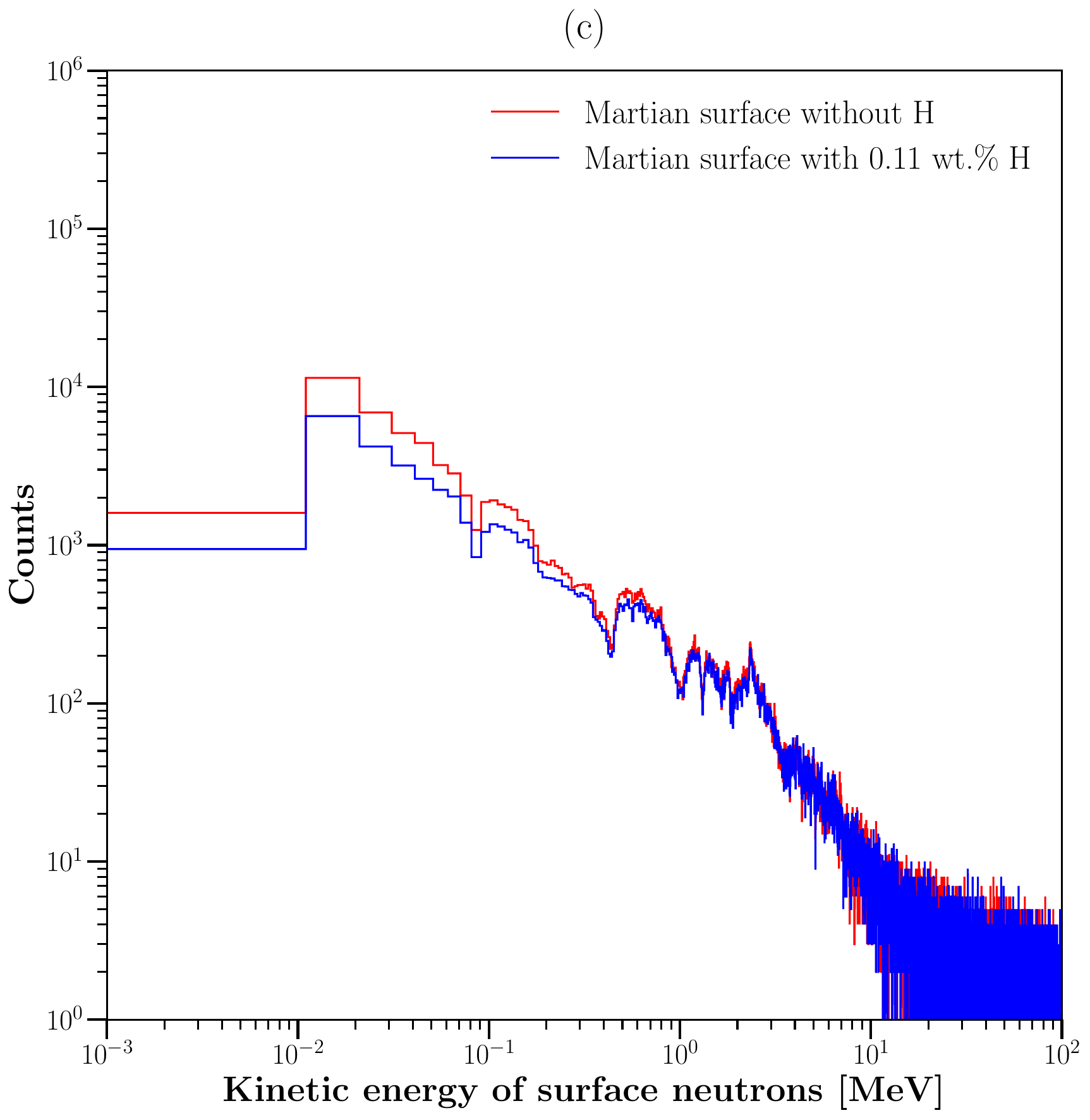}
\caption{Energy spectrum of albedo neutrons without/with 0.11 wt.\% hydrogen at the Martian surface (a) thermal albedo neutron spectrum, (b) epithermal albedo neutron spectrum, and (c) fast albedo neutron spectrum.}
\label{fig:surface_neutrons_divided}
\end{center}
\end{figure}

Concerning the neutrons and gammas produced within the atmosphere and arriving at the surface detector, Fig.~\ref{fig:atm_secondaries} depicts their kinetic energy spectra prior to any interaction with the regolith. As observed from Fig.~\ref{fig:atm_secondaries}(a), the kinetic energy spectrum of the atmospherically produced neutrons peaks below 1~MeV and falls off steeply toward higher energies, while Fig.~\ref{fig:atm_secondaries}(b) shows that the corresponding gamma-ray spectrum follows a similarly decreasing trend from its peak near 1~MeV, consistent with the soft bremsstrahlung and de-excitation photon component expected from an atmospheric hadronic cascade. These atmospherically produced particles constitute the downward-going source term that subsequently interacts with the regolith and drives the neutron and gamma production discussed next.
The profile of the generation depth for the secondary neutrons and gammas produced from the interaction of the atmospherically transported particles with the Martian regolith is obtained next. Fig.~\ref{fig:generation_depth}(a) presents the variation of the secondary neutron population with respect to generation depth in the absence and presence of 0.11 wt.\% hydrogen, while Fig.~\ref{fig:generation_depth}(b) shows the corresponding profile for the secondary gammas. As shown in Fig.~\ref{fig:generation_depth}, introducing 0.11 wt.\% hydrogen into the regolith has minimal impact on the generation depth of either the secondary neutrons or the secondary gammas, both of which peak at a depth of approximately 20--40~cm before decreasing monotonically with increasing depth in a manner that is essentially unchanged by the presence of hydrogen. This behavior indicates that hydrogen at this trace abundance does not appreciably alter the overall buildup and attenuation of the hadronic cascade within the regolith.
In order to assess the relative contribution of the atmosphere and the regolith to the overall neutron and gamma population reaching the Martian surface, Fig.~\ref{fig:surface_neutron_gamma} compares the atmospherically produced neutron and gamma spectra of Fig.~\ref{fig:atm_secondaries} with the total neutron and gamma spectra recorded at the surface detector without and with 0.11 wt.\% hydrogen. As shown in Fig.~\ref{fig:surface_neutron_gamma}(a), the total neutron population at the surface exceeds the purely atmospherically produced contribution by roughly two orders of magnitude below a few MeV, in both the dry and hydrogenated regolith cases, indicating that the Martian regolith constitutes the dominant source of the neutron population arriving at the surface, while the two contributions converge at higher energies, above approximately several tens of MeV, where the surface population becomes increasingly dominated by atmospherically produced neutrons that have not undergone further moderation within the regolith. A qualitatively similar picture emerges for the gamma-ray spectra in Fig.~\ref{fig:surface_neutron_gamma}(b), where the regolith-produced secondary gammas dominate the total surface gamma population at low and intermediate energies before converging with the atmospherically produced contribution at the highest energies considered. In both panels, the total neutron and gamma spectra obtained without and with 0.11 wt.\% hydrogen remain nearly indistinguishable from one another, consistent with the earlier finding that hydrogen at this trace abundance has minimal impact on the overall generation depth profile shown in Fig.~\ref{fig:generation_depth}. This comparison confirms that the Martian regolith, rather than the atmosphere alone, is the primary source of both the neutron and gamma radiation field at the surface, and it provides the context for the albedo neutron spectra discussed next.
In the next step, the albedo neutrons that escape from the Martian surface are gathered by utilizing the surface detector. The overall energy spectrum of the albedo neutrons acquired at the surface detector is divided into three energy ranges: thermal ($E\leq1$~eV), epithermal ($1~\text{eV}<E\leq1$~keV), and fast ($E>1$~keV). The thermal energy spectra of the albedo neutrons without and with 0.11 wt.\% hydrogen are exhibited in Fig.~\ref{fig:surface_neutrons_divided}(a). Although the presence of hydrogen is hardly observable in the generation depth of the secondary neutrons, as depicted in Fig.~\ref{fig:generation_depth}(a), it is revealed from Fig.~\ref{fig:surface_neutrons_divided}(a) that the introduction of 0.11 wt.\% hydrogen is traceable above the Martian surface, and the secondary neutrons undergo a non-negligible thermalization process before leaking out from the surface when hydrogen is present in the regolith: the number of thermal albedo neutrons is noticeably higher when hydrogen is present, particularly toward the lower end of the thermal range.
Secondly, the epithermal regime in the absence and presence of 0.11 wt.\% hydrogen is examined, the results of which are illustrated in Fig.~\ref{fig:surface_neutrons_divided}(b). Contrary to the thermal regime, the case without hydrogen exhibits more epithermal albedo neutrons across the entire epithermal energy range in comparison with the case that includes hydrogen. In opposition to the thermal regime, which shows a broad, parabolic-like trend along the thermal energy interval in Fig.~\ref{fig:surface_neutrons_divided}(a), the epithermal energy range leads to an approximately linearly decreasing trend on a log-log scale as demonstrated in Fig.~\ref{fig:surface_neutrons_divided}(b). Finally, Fig.~\ref{fig:surface_neutrons_divided}(c) shows the energy spectrum of the fast albedo neutrons in the absence and presence of 0.11 wt.\% hydrogen. It is observed that the case without hydrogen exhibits more fast albedo neutrons especially toward the lower end of the fast range, whereas the two cases begin to converge at higher energies, consistent with the fast neutron population there being dominated by the initial hadronic cascade rather than by moderation.
Taken together, the three panels of Fig.~\ref{fig:surface_neutrons_divided} reproduce the characteristic "epithermal deficit, thermal excess" pattern that constitutes the primary observable used to map subsurface water-equivalent hydrogen from orbit~\cite{maurice2011mars}, and they are qualitatively consistent with surface neutron spectra reconstructed from MSL/RAD and Odyssey/HEND measurements~\cite{kohler2014measurements,martinez2023unfolding}. It is thus concluded that the introduction of even a small amount of hydrogen into the Martian regolith is more influential on the moderation and escape of the albedo neutrons than on the proton transport or on the total neutron and gamma production depth profile, according to the present GEANT4 simulations.

\section{Conclusion}
\label{Conclusion}
In this study, the impact of introducing 0.11 wt.\% hydrogen on the neutron and gamma spectra at the Martian surface is investigated via the GEANT4 simulations. By employing a layered geometry consisting of the Martian atmosphere and a single-layer regolith in the absence and presence of 0.11 wt.\% hydrogen, it is observed that the range and energy deposition of protons within the regolith and the depth profile of the generated neutrons and gammas are only weakly sensitive to the trace hydrogen content considered here. In contrast, a significant thermalization and moderation of the secondary neutrons occurs before they leak out from the Martian surface if hydrogen is present in the regolith. The present findings reveal that the existence of 0.11 wt.\% hydrogen markedly increases the number of thermal albedo neutrons, while the epithermal albedo neutron population is consistently reduced across its entire energy range in the presence of hydrogen. This pattern is partially observed in the fast neutron regime as well, where the case without hydrogen exhibits more fast albedo neutrons toward the lower end of the fast range before the two cases converge at higher energies. It is demonstrated that the introduction of even a small amount of hydrogen substantially alters the albedo neutron spectra at the Martian surface, while leaving the proton transport and the overall secondary particle production depth profile comparatively unaffected by suggesting that neutron spectroscopy is a vital tool for detecting hydrogen, and possibly subsurface water ice, on Mars. These results are consistent with previous GEANT4- and PLANETOCOSMICS-based modeling of the Martian galactic cosmic-ray environment~\cite{matthiae2017radiation,matthia2017radiation,chen2022studies,guo2019implementation,charpentier2024aramis}, and with an analogous hydrogen-sensitivity effect previously reported for the lunar albedo neutron spectrum using a comparable GEANT4 methodology~\cite{topuz2024effect}, supporting the generality of hydrogen moderation as a diagnostic of subsurface hydration across airless and thin-atmosphere planetary bodies alike. Based on the present GEANT4 simulations, it is suggested that these insights might improve our understanding of Martian surface and subsurface composition and enhance the strategies for future robotic and human exploration of Mars.

\bibliographystyle{ieeetr}
\bibliography{Mars_neutron_gamma.bib} 
\end{document}